\documentstyle[preprint,prd,aps]{revtex}
\begin{document}
\draft
\preprint{}
\newcommand{\be}{\begin{equation}}
\newcommand{\ee}{\end{equation}}
\newcommand{\bea}{\begin{eqnarray}}
\newcommand{\eea}{\end{eqnarray}}
\title{A Measure on a Subspace of FRW Solutions\\ and
\\``The Flatness Problem" of Standard Cosmology}
\author{
H. T. Cho \footnote[1]{tkut174\%twnmoe10.bitnet@pucc.princeton.edu }}
\address{Tamkang University, Department of Physics,\\
Tamsui, Taipei, TAIWAN R.O.C.}
\vskip -.5 truein
\author{
R. Kantowski \footnote[2]{kantowski@phyast.nhn.uoknor.edu}}
\address{
University of Oklahoma, Department of Physics and Astronomy,\\
Norman, OK 73019, USA
}
\maketitle
\begin{abstract}
We use the metric on the space of gravity fields given by DeWitt to
construct a unique kinematic measure on  the space of FRW simple fluids and
show that when the mass parameter $\Omega$  is used as a coordinate
this measure is singular at $\Omega = 1$. This singularity, combined
with the time evolution of $\Omega$,  distorts distributions of  $\Omega$
values to be concentrated in the neighborhood of 1 at early times.
It is
a distorted distribution of  $\Omega$ values that sometimes misleads the
casual observer to conclude that  $\Omega$ must be exactly equal to 1.
\end{abstract}

\section{Introduction}
For decades now we have believed that general relativity determines the
dynamics of our universe and in fact one of the FRW
(Friedmann-Robertson-Walker) models closely approximates what we can
observe.  At the current epoch the dynamics of such a model is dominated
by inertia and the matter density (pressure now being insignificant and
the cosmological constant $\Lambda =0$). Our task has been
to find two observed numbers, e.g., $H_0$ the Hubble parameter and
$\Omega_0$ the mass density parameter [see (\ref{Einstein2}) and
(\ref{Omega0}) for definitions], and hence  to tie down completely
the global structure
of the universe, as well as where we are in its time development. The
accepted  value of
the Hubble parameter is somewhere between 40-90 km/s/Mpc, depending on
how it is estimated \cite{FM}. Advocates for one extreme value or the
other are
not supported by some fundamental principle which makes their value more
appealing.
The same is not true for the other parameter $\Omega_0$. Its
accepted\footnote{ The dynamical values obtained by \cite{KN,DA}
are much larger. We chose to quote and use values obtained from more
established methods simply to make our point about the problems
encountered when  $\Omega$ is used as a coordinate.} value from
observation is between 0.01-0.2 (luminous - dynamical mass) \cite{KE,DM}
with the
frequently advocated value being 1. When $\Omega =1$
the Universe is on the verge of being closed even though the spatial
sections are flat.  If currently  $\Omega \approx 1$ then at earlier
times (as argued below) $\Omega \rightarrow 1$ and as can be seen in
(\ref{Einstein2}) the spatial curvature of the universe ($k/R^2$) had
negligible effect on its early dynamics. This is referred to as the
``flatness problem" of standard cosmology. The advent of Inflation has
added fervor to the
debate because, in addition to solving some long standing problems of
cosmology  (in particular the horizon problem), it would guarantee the
almost sanctified
value of $\Omega =1$. Sometimes when listening to advocates for inflation
the
audience is misled to think that the ``flatness problem" implies that
$\Omega$ is exactly equal to 1 in the early universe and hence inflation
must be correct.
The failure to now observe the value $\Omega_0 =1$, becomes the devotee's
``$\Omega$ problem" or equivalently an additional missing mass problem.

The argument goes something like the following \cite{DR,LA,GA}: If at
present, the value
$\Omega = \Omega_0 = 1 - \delta_0$, then at earlier times
$\Omega ={\Omega_0(1+z) / (1+\Omega_0 z)}\approx 1-\delta_0/ z$.  This
value gets
closer and closer to 1  as you choose earlier and earlier  times,
e.g., when the effects of pressure on the expansion of the universe are
no longer negligible
$ z_{R} \approx 10^{4}$ and $\delta_{R}\approx\delta_0\times 10^{-4}$.
Before this period when radiation is dominating the expansion, $\Omega$
is approaching 1 even faster,
$\Omega =\Omega_{R}(1+z)^2/(1+2\Omega_{R} z+\Omega_{R}z^2)
\approx 1 -\delta_{R}/z^2$. At the time of nucleosynthesis where
$z\approx 10^{10}= 10^6$ relative to $z_{R}$ we are sure of our $\approx 1$
MeV physics and we have $\Omega = 1-\delta_0\times 10^{-16}$. If the
Universe would have undershot 1 by some
reasonable value such as $10^{-5}$  at this early epoch then there wouldn't
be much around now, including us; and if the universe had overshot 1 by such
a reasonable value then it would have collapsed long ago. The misleading
conclusion drawn from such or similar arguments is that $\Omega$ must
exactly equal one, after all, ``How
could it be so close and not be 1?". This conclusion is based on an
unstated assumption that at some early epoch our value of $\Omega$
should have been chosen from some possible set of values (by either a
classical or quantum mechanical process) of which $\Omega=1$ was no
more likely than any other value (see \cite{GE} for a discussion of
initial data). By introducing a measure on a subspace of FRW  solutions
we expose
 $\Omega$ as  the problem, i.e., that probability distributions will be
skewed towards  $\Omega =1$, and that if a ``better" coordinate is used
the flatness problem clearly doesn't imply $\Omega =1$. In Sec. 2 we
introduce a ``better" coordinate called $C$ and in Sec. 3 we introduce
the essentially unique measure (the kinematic measure) on the space of
solutions and express it in both the ``good" coordinate $C$ and the not
so good coordinate  $\Omega$. In Sec. 4  we make the point about $\Omega$
being a ``bad" coordinate by following a hypothetical distribution to
larger and larger redshifts. We also conjecture the relationship of the
 kinematic measure proposed here to the dynamical measure proposed by
Henneaux \cite{HM} and Gibbons et al. \cite{GG}.

\section{A Coordinate for Simple Perfect Fluid FRW Solutions}

The Robertson-Walker metrics can be found in every book on cosmology,
e.g., see
\cite{RW},
\be
d s^2 = c^2 d t^2 -R(t)^2[ {d r^2\over 1-k r^2} +r^2( \sin^2\theta
d \phi^2 +d \theta^2)]\,,
\label{RW}
\ee
where $ k = -1, 0, 1$ and $R(t)$ is arbitrary.  For simple perfect fluid
solutions ($ p = (\gamma-1)\rho c^2$) of the standard theory, $R(t)$
is determined by the Einstein equations which reduce to:
\be
{8\pi G\over 3 c^2}\rho R^{3\gamma} =   \text{constant}
\equiv C^{3\gamma-2}\ ,
\label{Einstein1}
\ee
and
\be
{H^2\over c^2} = {8\pi G\over 3 c^2}\rho -{k\over R^2}={1\over
R^2}[\left({C\over R}\right)^{3\gamma-2}-k] \ ,
\label{Einstein2}
\ee
where $H\equiv \dot{R}/R$. The constant in (\ref{Einstein1}) has
been written in terms of another constant $C$ whose units are the
same as those of $R$. The one parameter family of solutions  $R(t,C)$
is given by
integrating (\ref{Einstein2}),
\be
\int dR \left[\left({C\over R}\right)^{3\gamma-2}-k \right]^{-1/2 }
= c\int dt.
\label{Friedmann}
\ee

For the spatially flat $k=0$ case, $C$ can be scaled to any desired
value by scaling the $r$ coordinate and hence only one such solution
exists. The same is not true for the spatially curved $k = \pm 1$ solutions;
$C$ remains as the single parameter ($0 < C < \infty$) distinguishing
between possible models. For the closed FRW models $C$ is clearly the
maximum value of $R$. The current value of $C$ ($\gamma =1$ for
pressure =0) corresponding to the above observed range of small
$\Omega_0$ values is $C_0 =(0.01 - 0.3) c/H_0$. Once $C$ is fixed
another parameter (e.g., $t_0$ or $H_0$) must be given to fix our epoch.
Giving the Hubble parameter $H_0 = H(t_0=t_{now})$ is equivalent to
giving the
current critical mass density $\rho_c$ of the universe,
\be
\rho_c = {3H^2_o\over 8\pi G} \ .
\label{rhoc}
\ee
The mass density parameter $\Omega_0$ is normally used as a label for
solutions rather than the $C$ introduced above. It is defined in terms
of the current
mass density $\rho_0$ and its critical value,
\be
\Omega_0 \equiv {\rho_0\over \rho_c} \ .
\label{Omega0}
\ee
In what follows we use $C$ and $\Omega_0$ as two different
parameterizations of the above set of gravity fields.

\section{The Invariant Measure on the Space of FRW Solutions}

To statistically weight a set of possible fields $\{ \phi^i \}$,
two structures must be given: (i) a measure (e.g., a volume element)
on the space of fields  and (ii) a scalar function normalized with the
given measure. For many fields (including the metric fields
$ g_{\alpha\beta}(x)$ of gravity) the only known measure is proportional
to the volume element of some field metric $G_{ij}(\phi)$ on the
space of fields,
\be
ds^2 = G_{ij} d\phi_{\perp}^i d\phi_{\perp}^j = G^{\perp}_{ij}
d\phi^i d\phi^j \ . \label{Gmetric}
\ee
The parallel  projection, $d\phi_{\parallel}^i = P^i_{\parallel j}
d\phi^j$, selects the gauge dependent part of the difference of two
neighboring fields and the perpendicular projection
$ d\phi_{\perp}^i=(\delta^i_j-P^i_{\parallel j}) d\phi^j$ selects
the part orthogonal to all possible gauge transformations,
\be
G_{ij}P^i_{\parallel k} \left(\delta^j_l-P^j_{\parallel l}\right) = 0\ ,
\ee
giving
\be
 G^{\perp}_{ij} \equiv  G_{ij}-G_{kl} P^k_{\parallel i}P^l_{\parallel j}\ .
\label {Gperp}
\ee
The distance between two gauge equivalent fields, computed using
(\ref{Gperp}), clearly vanishes. Other measures can be defined if
the set of fields is restricted by some dynamical theory,  e.g. a phase
space volume can be defined when the dynamics is canonically described.
For the non-dynamically restricted
metric fields a unique field-metric exists and is commonly used when
performing a path intergal quantization of gravity \cite{BA}. It was
first given by DeWitt \cite{DB1} but its absolutely essential role was
made clear when Vilkovisky developed the current effective action theory
\cite{VG,DB2}.
We fix the differential manifold and write the field in a given
coordinate patch as
\be
\phi^i = g_{\alpha\beta}(x) \quad\quad   ( i = \{\alpha,\beta,x\} ) \ .
\ee
The field-space metric of DeWitt \cite{DB1} to be used in (\ref{Gmetric})
to give the distance between two neighboring metrics is
\be
G_{ij} = \sqrt{|{\rm det}\ g|}{1\over 4}\left[g^{\alpha\mu} g^{\beta\nu}+
g^{\alpha\nu} g^{\beta\mu}-a\ g^{\alpha\beta}
g^{\mu\nu}\right]_x \delta^4(x-y)\ ,
\label{metric}
\ee
where $a$ is an arbitrary unitless constant ($\ne 1/2$).
This metric is commonly used in path integral versions of quantum
gravity; however, it is a purely classical structure and it is only in
that context that we use it here.

For metric fields the gauge group is the set of active coordinate
transformations (i.e. the diffeomorphism group) and
the difference between two neighboring fields is decomposed into a
part attributable to an active coordinate change and a part which
is not, i.e. a part perpendicular to all possible coordinate changes
(see Appendix),
\be
 \delta \phi^i = \delta g_{\alpha\beta}(x) =
\delta g_{\parallel\alpha\beta}(x)  + \delta g_{\perp\alpha\beta}(x) \ .
\ee
Here $\delta g_{\parallel\alpha\beta}(x)
=\nabla_{\alpha}\delta\xi_{\beta}+\nabla_{\beta}\delta\xi_{\alpha}$
is generated by some small coordinate shift $x^{\alpha}
\rightarrow x^{\alpha} + \delta\xi^{\alpha}(x)$.
The metric as given by  (\ref{metric}) is unique (up to the
parameter $a$) provided  that $G_{ij}$ is assumed to be local
( i.e. $ \propto \delta^4(x-y)$), assumed not to depend on the
metric's curvature (i.e. not to depend on derivatives of
$g_{\alpha\beta}$), and assumed to be invariant under gauge
transformations. In equations (\ref{measure}) and (\ref{normprob})
we will see that the value of the arbitrary parameter $a$ doesn't affect
a normalized probability distribution on the FRW subspace studied here.
Equation (\ref{Gmetric}), evaluated using  (\ref{metric}), should be
thought of as giving the intrinsic (i.e., coordinate independent)
geometrical distance between two metrics $ g_{\alpha\beta}(x)$ and
$g_{\alpha\beta}(x) + \delta g_{\alpha\beta}(x)$ defined on the same
manifold.
The induced natural (kinematic) measure associated with a set of metric
fields is simply proportional to the volume of a neighboring set of
fields, i.e., $\propto  \text{det}|G^{\perp}_{ij}|$.
The above metric (\ref{metric}) on all metric fields will induce a metric
on any subspace of fields; in particular it will induce a metric $G(C)$ on
the $\gamma = $ fixed subspaces of $k=\pm 1$
perfect fluid FRW solutions,
\be
ds^2 = G^{\perp}_{ij} d\phi^i d\phi^j = G(C) dC dC \ .
\label{fieldmetric2}
\ee
The $k = 0$ solution is only a point in the field space. The induced
natural measure on the open (closed) simple fluid
solutions is  $\propto \sqrt{G(C)}\ dC$.
To compute it we rewrite (\ref{RW})  replacing $t$ by a new variable
$\chi\equiv R/C$
\be
dt = {dR\over H R} ={C d\chi \over c\sqrt{\chi^{2-3\gamma}-k}}\ .
\ee
The form of the metric is now
\be
ds^2_C = C^2\left\{ {d^2\chi\over \chi^{2-3\gamma}-k} -
\chi^2\left[ {d r^2\over 1-k r^2} +r^2( \sin^2\theta
d \phi^2 +d \theta^2)\right] \right\}
=C^2 ds^2_{C=1}\ .
\label{FRW}
\ee
The range of the new coordinate $\chi$ is $ 0\le \chi
< \chi_{\text max}$ where $ \chi_{\text max} = \infty$
for $k=-1$ and  $ \chi_{\text max} = 1$ for $k=1$.
The difference in two neighboring metric fields of fixed
$\gamma$ becomes
\be
\delta g_{\alpha\beta}(C;\chi,r,\theta,\phi) =
2C\delta C\ g_{\alpha\beta}(C=1;\chi,r,\theta,\phi)\ ,
\ee
written symbolically as
\be
\delta\phi^i_C =2C\delta C \phi^i_{C=1}\ ,
\ee
and giving from (\ref{fieldmetric2}) an induced metric
\be
G(C)dCdC= 4 C^2 G^{\perp}_{ij}(C)\phi^i_{C=1}
\phi^j_{C=1} dCdC\ .
\label{GC}
\ee
{}From (\ref{metric}) and (\ref{FRW}) it is clear that in 4-dimensions
$G_{ij}(C)=G_{ij}(C=1)$, i.e. that the field metric when evaluated at
any of the $\gamma$ = fixed simple fluids doesn't depend on $C$. In
the Appendix we show that the same is true for $ G^{\perp}_{ij}(C)$
[see  (\ref{Gperp=con})], consequently giving the measure as a
simple function of $C$,
\be
\sqrt{G(C)} = {\rm constant}\times C\ .
\ee
The potentially devastating divergence that occurs (constant
$\rightarrow \infty$) for the infinite open models is harmless
here because we are keeping the equation of state fixed and a
normalization of probability removes the constant.
The parameters $H_0$ and $\Omega_0$ rather than $C$ are
ordinarily used to label the FRW solutions. Of these two
parameters $H_0$ is fixed at its
current value and  $\Omega_0$ is used as the free parameter.
Eliminating $R$ between equations (\ref{Einstein1}) and
(\ref{Einstein2}) gives
\be
C= {c\over H_0}\left({-k\over 1-\Omega_0}
\right)^{{3\gamma\over 2(3\gamma-2)}} \Omega_0^{1\over 3\gamma-2}\ ,
\label{C}
\ee
which implies
\be
dC = -k{c\over H_0} \left({-k\over 1-\Omega_0}\right)^{9\gamma-
4\over 2(3\gamma-2)}\left( {1\over 3\gamma -2} + {\Omega_0\over 2}\right)
\Omega_0^{3-3\gamma \over 3\gamma -2}\ d\Omega_0\ .
\label{dC}
\ee
The measure as a function of $\Omega_0$ becomes
\bea
\sqrt{G(C)}\ dC &=& \sqrt{G(\Omega_0)}\ d\Omega_0\ \nonumber\\
&=& {\rm constant}\times C dC\  \nonumber\\
&=& {\rm constant}\times (-k)\left({c\over H_0}\right)^2
\left({-k\over 1-\Omega_0}\right)^{2(3\gamma-1)\over3\gamma-2}
\left({1\over 3\gamma -2} +{\Omega_0\over 2}\right)
\Omega_0^{4-3\gamma\over 3\gamma-2}\ d\Omega_0\ ,
\label{measure}
\eea
and clearly diverges on any neighborhood of $\Omega_0 = 1$ when
$\gamma > 2/3$. This expression is the distance between two neighboring
universes whose coordinates are  $C$  and $C+dC$. In the second form the
distance is evaluated by comparing the values of $\Omega$ for these to
universes when their Hubble parameters are the same (both $=H_0$).
\section{Conclusions and discussion}

We have not proposed  any dynamical  mechanism to determine the
distribution of possible FRW universes. We only argue that $\Omega$
is not the best coordinate to use for a label if you wish to consider
earlier and earlier times. The  only natural measure on the space of
FRW polytropic solutions is singular at $\Omega = 1$ and (as seen below)
every neighborhood of 1 shrinks to 1 at early times.
If the parameter C is used, its value is
well behaved in the currently observed negligible pressure  domain
 $C_0 < \infty$, [see Eqn. (\ref{C}) with $\gamma =1$],
 and that this value remains constant all the way back to a period when
radiation
rather than $p=0$ models describe the dynamics ($z_{R}\approx 10^4$).
Matching boundary conditions ($R,\dot R,\,{\rm and}\, \rho$) at this
redshift  where the equation of state changes to $\gamma =4/3$ requires
a  decrease in the value of the constant $C$,
\be
C_{R}=\sqrt{R_0C_0/(1+z_{R})} = {c\over H_0}(0.001 \rightarrow 0.006).
\ee
This constant value persists as far back as the equation of state
($\gamma = 4/3$) remains valid, e.g. to the Inflation period.

If we assume this observed value exists by choice among some
normalized set of possible values,  a length scale $L$ must
exist for the distribution function $P(C^2/L^2)$,
\be
\int_0^{\infty}P(C^2/L^2) d(C^2/L^2) = 1\ ,
\label{normprob}
\ee
and we can immediately see the true flatness problem: why is
$L \approx C_{R} \approx 10^{59} \times L_{Planck}$?
If this distribution was determined at the time of transition
from quantum gravity to classical gravity when the only length
around was the Planck length ($ L_{Planck} = 1.6 \times 10^{-33}$ cm),
what inflated it by a factor of $10^{59}$\ ? One of the current forms
of Inflation is commonly assumed to have done so; however, \cite{HS1,DC}
argues that $\Omega$ could be $\approx 1$ without inflation.
The actual form of $P(C^2/L^2)$ is of course not known but its origin
must be determined by the probability of having sources of gravity which
produce a given gravity field , i.e. a given  $C$. For illustrative purposes
we pick a simple normalized example,
\be
P(C^2/L^2) = \exp{(-C^2/L^2)}\ .
\label{exp}
\ee
Using (\ref{C}) and (\ref{dC}) with $z =0$ replaced by $z_R$ and
$\gamma =4/3$ along with the redshift dependence of $\Omega$
computed from (\ref{Einstein2}), (\ref{rhoc}), and (\ref{Omega0}), i.e.,
\be
\Omega_R = {1\over 1+(1/\Omega -1)(1+z)^{3\gamma-2}},
\label{Omega}
\ee
we can look at the distribution of possible $\Omega $ values at
early times by writing
\be
P(\Omega, z) d\Omega = P(C^2/L^2) d (C^2/L^2).
\label{prob}
\ee
In (\ref{Omega}) $z = 0$ is at the end of the radiation phase where the mass
parameter is $\Omega_R$.
What is found (e.g., see Fig. 1) is a distribution rapidly being
squeezed (as $z$ increases) to a peak just less than $\Omega =1$.
The narrowing peak follows the implicit solution $\Omega(z)$ of
equation (\ref{Omega}). It is cut off on the left by the fact that
the distribution is normalized [e.g. by the exponential  in (\ref{exp})]
and on the right by the singularity in the measure (\ref{measure}).
The maximum in the probability curve is going up as $(1+z)^2$, the
width is shrinking as  $(1+z)^{-2}$, and the difference $1-\Omega$ is
decreasing as  $(1+z)^{-2}$. It is this narrow, extremely high peak
being squeezed to $\Omega =1$ that frequently misleads a
casual observer to think that  $\Omega$ must be ``fine tuned" to 1.
In our simple example the probability density actually vanishes at
$\Omega =1$.

Alternatively you could argue that by forcing a uniform distribution
of $C^2$ (i.e., $L \rightarrow \infty$), you force $\Omega \rightarrow 1$
as the only value allowed for $\Omega$. Without a  scale for $C$ the only
choices are $L =0$ or $L=\infty$ which correspond to $\Omega \rightarrow 0$
and $\Omega \rightarrow 1$
respectively.

Other measures on the space of FRW solutions have been proposed in
conjunction with classical \cite{HM,GG,HS1} or quantum \cite{NJ,HS2}
dynamical theories.
The $\gamma =2$ case given here can be directly compared with the
massless scalar field case of Gibbons et al. \cite{GG}, see equation
(3.15). Here the gravity field space is clearly 1 dimensional
($C$ is 1 parameter), but there the Henneaux, Gibbons, Hawking,
and Stewart measure is for a 2-dimensional initial data space.
The extra dimension appearing in the dynamical measure comes from
the initial data for the scalar field $\phi$. The value of the scalar
field doesn't effect the gravity field (only its rate of change does)
and,
not surprisingly, their measure is of the form
\be
d\mu = constant\times dC^2\wedge d\phi\ ,
\ee
when our $C$ coordinate is used. In the form given by
Gibbons et al. \cite{GG} the measure is of the form of our
equation (\ref{measure})$\wedge d\phi$ (their coordinate
$y\equiv H_0 \sqrt{\Omega_0}$). Integrating over the $\phi$ initial
data gives a uniform distribution in $C^2$, i.e. $L \rightarrow \infty$ and
$\Omega   \rightarrow 1$. The origin of their result is clear. The gravity
field part, $\sqrt{G(C)}\ dC$ (the kinematic measure as we call it) is as
we say it inevitably must be and the massless scalar field, having no
intrinsic scale and having had its initial (dynamical) value uniformly
distributed, cannot select any one $C$ over another, i.e., $P(C^2/L^2)$
is constant. Normalization forces this constant to zero and
selects the divergent point $\Omega \rightarrow 1$ as the only possible
configuration.
For other more complicated cases we expect similar agreement between the
unique  kinematic measure we propose and dynamical probability distribution
coming from the canonical phase space measure proposed by
Henneaux, Gibbons, Hawking, and Stewart. For more complicated
cases this agreement is likely to occur only when the parameter
$a=1$ in (\ref{metric}). This is because the $a=1$ metric appears
in  the kinetic energy term for background field expansions and is
hence built into
any dynamical theory containing conventional GR.

Our objective here has been limited to evaluating the unique kinematic
measure induced on the configuration space of a limited set of gravity
fields. We have found that it is not well behaved as
$\Omega \rightarrow 1$. In addition we hope we have convinced
the reader of two things:

\noindent 1)\quad
That in the absence of knowing  the true distribution function
of expected values of $\Omega_0$  or the dynamical mechanism
that produces the distribution, one should use the measure
given here simply because of its uniqueness. If the probability
of producing a given gravity field  by the set of all sources
were to be known, it would appear as the function $P(C^2/L^2)$,
normalized with this measure
as in (\ref{normprob}).

\noindent 2)\quad
That the assumption  $\Omega_0 \approx 1$ implies $\Omega = 1$ is
 based on an unstated assumption that the distribution of possible
values of
$\Omega$ is relatively flat at $\Omega =1$. If it were well behaved
at 1, finding a value differing from 1 by 10$^{-16}$ or less would
be deemed significant. It would imply that some additional mechanism
beyond conventional dynamics and probabilities produced the  observed
early values of $\Omega\approx 1$, e.g., Inflation
might have driven $\Omega_0$ to this value.
However, we know that  $\Omega$ is not a good coordinate to use because
a divergence in the  measure  will amplify the probability distribution
as $\Omega \rightarrow 1$. Consequently finding an early value near
$\Omega=1$ might be quite likely
even if the probability of finding a value of $\Omega=1$ was zero.

Finally, we know the production of a distribution of $\Omega$'s is
one thing, but observing various values  is another. Only those
universes or
parts of ``the universe"  having a limited range
of $H_0$ and $\Omega_0$  values would likely produce civilizations
such as ours
asking such questions. This selection effect cannot be denied. However,
it may or may not have distorted the original distribution. In any event,
this selected  distribution is likely to include only universes  where
$\gamma\ge 1$  for a significant recent history and for all of these,
$\Omega$ approaches 1 at earlier times.

\section{Acknowledgements}
The authors thank D. Branch for suggesting improvements to this
manuscript and D. Page for finding a logical error in the first version
as well as motivating several improvements.
 This work was supported
by the Department of Energy, and the Southern Association for High
Energy
Physics (SAHEP) funded by the Texas National Research Laboratory
Commission (TNRLC). H. T. Cho is supported by grant \#  NSC83-0208-M-032-034.

\section{Appendix}

What is referred to as the gauge group for metric fields on a fixed
differentiable manifold is actually the group of diffeomorphisms of
that manifold. All metrics are identified as equivalent that can be actively
transformed into one another. For `infinitesimal'  transformations these
look like $x^{\alpha} \rightarrow x^{\alpha} + \delta\xi^{\alpha}(x)$ which
change
the metric by
\be
g_{\alpha\beta}(x) \rightarrow g_{\alpha\beta}(x) +
\nabla_{\alpha}\delta\xi_{\beta}+\nabla_{\beta}\delta\xi_{\alpha}\ ,
\ee
and which are generically written as
\be
\phi^i \rightarrow \phi^i + Q^i_{\sigma}\delta\xi^{\sigma}\ ,
\ee
where
\be
Q^i_{\sigma}\delta\xi^{\sigma} = \int d^4y
\left\{ g_{\alpha\gamma}\nabla_{\beta}+g_{\beta\gamma}
\nabla_{\beta}\right\}_x\delta^4(x-y) \delta\xi^{\gamma}(y)\ ,
\label{Q}
\ee
i.e., where $(i =\{\alpha,\beta,x\})$ and $(\sigma =\{\gamma,y\})$,
repeated
discrete indices are summed over, and repeated continuous indices are
integrated over. The metric components in the gauge directions are defined by
\be
N_{\sigma\rho} = G_{ij}Q^i_{\sigma}Q^j_{\rho} = -
\sqrt{-g}\left\{ g_{\sigma\rho}\Box + \nabla_{\rho}
\nabla_{\sigma}-a \nabla_{\sigma}\nabla_{\rho}\right\}_y\delta^4(y-z)\ ,
\label{N}
\ee
and are seen to form a local differential operator whose
inverse $N^{\sigma\rho}$ is  consequently a non-local Green's function,
\be
N^{\sigma\lambda} N_{\lambda\rho} = \delta^{\sigma}_{\rho}\delta^4(y-z)\ .
\label{N-1def}
\ee
The relevant quantity needed for computing $G^{\perp}_{ij}$ is the parallel
projection operator
\be
P^i_{\parallel j} = Q^i_{\sigma}N^{\sigma\rho}Q^k_{\rho}G_{kj}\ ,
\ee
and is non-local because of the $N^{\sigma\rho}$ term.
The perpendicular part of the field metric needed is consequently
\be
G^{\perp}_{ij} = G_{ij} - G_{ik}Q^k_{\sigma}N^{\sigma\rho}Q^l_{\rho}G_{lj}\ .
\label{Gperp2}
\ee
What we wish to show is that $G^{\perp}_{ij}$  like  $G_{ij}$ (as we have
already pointed out in the paragraph after eqn. (\ref{GC})) when evaluated
at (\ref{FRW}) is independent of $C$. From  (\ref{Q}) we see
$Q^i_{\sigma}(C) = C^2 Q^i_{\sigma}(C=1)$,
and from (\ref{N}), $N_{\sigma\rho}(C) =  C^4 N_{\sigma\rho}(C=1)$.
{}From (\ref{N-1def}) we see
$N^{\sigma\rho}(C) = C^{-4} N^{\sigma\rho}(C=1)$,
and consequently from (\ref{Gperp2})  we have the desired result
\be
G^{\perp}_{ij}(C) = G^{\perp}_{ij}(C=1)\ .
\label{Gperp=con}
\ee

\vskip .25 truein
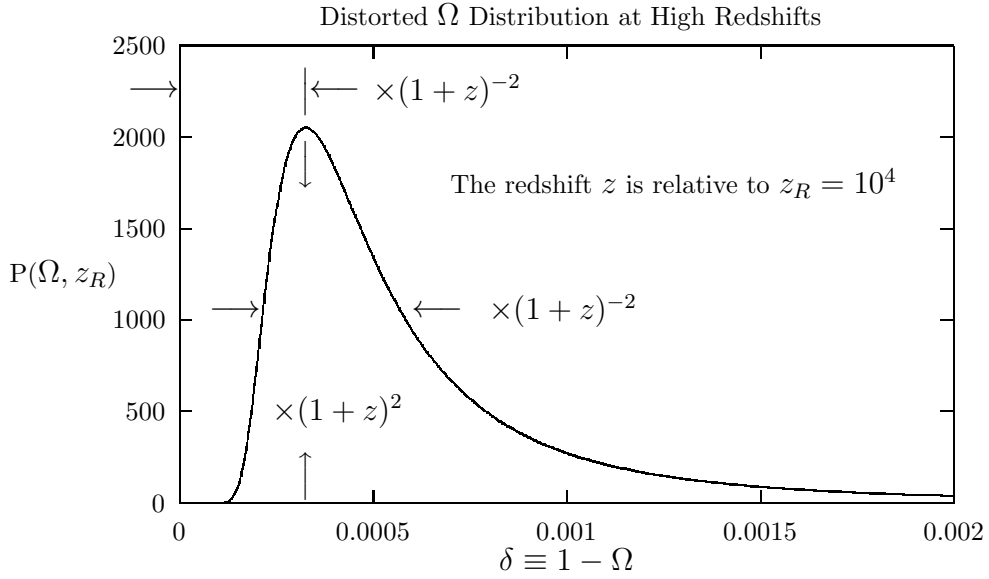
\begin{figure}
\begin{center}
\vskip 1.0 truein

\setlength{\unitlength}{0.240900pt}
\ifx\plotpoint\undefined\newsavebox{\plotpoint}\fi
\sbox{\plotpoint}{\rule[-0.200pt]{0.400pt}{0.400pt}}%
\begin{picture}(1500,900)(0,0)
\font\gnuplot=cmr10 at 10pt
\gnuplot
\sbox{\plotpoint}{\rule[-0.200pt]{0.400pt}{0.400pt}}%
\put(220.0,113.0){\rule[-0.200pt]{292.934pt}{0.400pt}}
\put(220.0,113.0){\rule[-0.200pt]{0.400pt}{173.207pt}}
\put(220.0,113.0){\rule[-0.200pt]{4.818pt}{0.400pt}}
\put(198,113){\makebox(0,0)[r]{0}}
\put(1416.0,113.0){\rule[-0.200pt]{4.818pt}{0.400pt}}
\put(220.0,257.0){\rule[-0.200pt]{4.818pt}{0.400pt}}
\put(198,257){\makebox(0,0)[r]{500}}
\put(1416.0,257.0){\rule[-0.200pt]{4.818pt}{0.400pt}}
\put(220.0,401.0){\rule[-0.200pt]{4.818pt}{0.400pt}}
\put(198,401){\makebox(0,0)[r]{1000}}
\put(1416.0,401.0){\rule[-0.200pt]{4.818pt}{0.400pt}}
\put(220.0,544.0){\rule[-0.200pt]{4.818pt}{0.400pt}}
\put(198,544){\makebox(0,0)[r]{1500}}
\put(1416.0,544.0){\rule[-0.200pt]{4.818pt}{0.400pt}}
\put(220.0,688.0){\rule[-0.200pt]{4.818pt}{0.400pt}}
\put(198,688){\makebox(0,0)[r]{2000}}
\put(1416.0,688.0){\rule[-0.200pt]{4.818pt}{0.400pt}}
\put(220.0,832.0){\rule[-0.200pt]{4.818pt}{0.400pt}}
\put(198,832){\makebox(0,0)[r]{2500}}
\put(1416.0,832.0){\rule[-0.200pt]{4.818pt}{0.400pt}}
\put(220.0,113.0){\rule[-0.200pt]{0.400pt}{4.818pt}}
\put(220,68){\makebox(0,0){0}}
\put(220.0,812.0){\rule[-0.200pt]{0.400pt}{4.818pt}}
\put(524.0,113.0){\rule[-0.200pt]{0.400pt}{4.818pt}}
\put(524,68){\makebox(0,0){0.0005}}
\put(524.0,812.0){\rule[-0.200pt]{0.400pt}{4.818pt}}
\put(828.0,113.0){\rule[-0.200pt]{0.400pt}{4.818pt}}
\put(828,68){\makebox(0,0){0.001}}
\put(828.0,812.0){\rule[-0.200pt]{0.400pt}{4.818pt}}
\put(1132.0,113.0){\rule[-0.200pt]{0.400pt}{4.818pt}}
\put(1132,68){\makebox(0,0){0.0015}}
\put(1132.0,812.0){\rule[-0.200pt]{0.400pt}{4.818pt}}
\put(1436.0,113.0){\rule[-0.200pt]{0.400pt}{4.818pt}}
\put(1436,68){\makebox(0,0){0.002}}
\put(1436.0,812.0){\rule[-0.200pt]{0.400pt}{4.818pt}}
\put(220.0,113.0){\rule[-0.200pt]{292.934pt}{0.400pt}}
\put(1436.0,113.0){\rule[-0.200pt]{0.400pt}{173.207pt}}
\put(220.0,832.0){\rule[-0.200pt]{292.934pt}{0.400pt}}
\put(45,472){\makebox(0,0){P($\Omega,z_R$) }}
\put(828,23){\makebox(0,0){$\delta \equiv 1 - \Omega$}}
\put(828,877){\makebox(0,0){ Distorted $\Omega$
Distribution at High Redshifts}}
\put(646,616){\makebox(0,0)[l]{The redshift $z$
is relative to $z_R = 10^4$ }}
\put(585,415){\makebox(0,0)[l]{$\longleftarrow$}}
\put(348,415){\makebox(0,0)[r]{$\longrightarrow$}}
\put(418,645){\makebox(0,0){$\Big\downarrow$}}
\put(418,760){\makebox(0,0){$\Big\vert$}}
\put(418,156){\makebox(0,0){$\Big\uparrow$}}
\put(424,760){\makebox(0,0)[l]{$\longleftarrow$}}
\put(220,760){\makebox(0,0)[r]{$\longrightarrow$}}
\put(524,760){\makebox(0,0)[l]{$\times (1+z)^{-2}$}}
\put(706,415){\makebox(0,0)[l]{$\times (1+z)^{-2}$}}
\put(366,257){\makebox(0,0)[l]{$\times (1+z)^{2}$}}
\put(220.0,113.0){\rule[-0.200pt]{0.400pt}{173.207pt}}
\put(232,113){\usebox{\plotpoint}}
\put(289,112.67){\rule{0.964pt}{0.400pt}}
\multiput(289.00,112.17)(2.000,1.000){2}{\rule{0.482pt}{0.400pt}}
\put(293,113.67){\rule{0.964pt}{0.400pt}}
\multiput(293.00,113.17)(2.000,1.000){2}{\rule{0.482pt}{0.400pt}}
\multiput(297.00,115.61)(0.685,0.447){3}{\rule{0.633pt}{0.108pt}}
\multiput(297.00,114.17)(2.685,3.000){2}{\rule{0.317pt}{0.400pt}}
\multiput(301.60,118.00)(0.468,0.627){5}{\rule{0.113pt}{0.600pt}}
\multiput(300.17,118.00)(4.000,3.755){2}{\rule{0.400pt}{0.300pt}}
\multiput(305.60,123.00)(0.468,1.066){5}{\rule{0.113pt}{0.900pt}}
\multiput(304.17,123.00)(4.000,6.132){2}{\rule{0.400pt}{0.450pt}}
\multiput(309.59,131.00)(0.477,1.155){7}{\rule{0.115pt}{0.980pt}}
\multiput(308.17,131.00)(5.000,8.966){2}{\rule{0.400pt}{0.490pt}}
\multiput(314.60,142.00)(0.468,2.236){5}{\rule{0.113pt}{1.700pt}}
\multiput(313.17,142.00)(4.000,12.472){2}{\rule{0.400pt}{0.850pt}}
\multiput(318.60,158.00)(0.468,2.674){5}{\rule{0.113pt}{2.000pt}}
\multiput(317.17,158.00)(4.000,14.849){2}{\rule{0.400pt}{1.000pt}}
\multiput(322.60,177.00)(0.468,3.406){5}{\rule{0.113pt}{2.500pt}}
\multiput(321.17,177.00)(4.000,18.811){2}{\rule{0.400pt}{1.250pt}}
\multiput(326.60,201.00)(0.468,3.990){5}{\rule{0.113pt}{2.900pt}}
\multiput(325.17,201.00)(4.000,21.981){2}{\rule{0.400pt}{1.450pt}}
\multiput(330.60,229.00)(0.468,4.283){5}{\rule{0.113pt}{3.100pt}}
\multiput(329.17,229.00)(4.000,23.566){2}{\rule{0.400pt}{1.550pt}}
\multiput(334.60,259.00)(0.468,4.868){5}{\rule{0.113pt}{3.500pt}}
\multiput(333.17,259.00)(4.000,26.736){2}{\rule{0.400pt}{1.750pt}}
\multiput(338.60,293.00)(0.468,4.868){5}{\rule{0.113pt}{3.500pt}}
\multiput(337.17,293.00)(4.000,26.736){2}{\rule{0.400pt}{1.750pt}}
\multiput(342.60,327.00)(0.468,5.160){5}{\rule{0.113pt}{3.700pt}}
\multiput(341.17,327.00)(4.000,28.320){2}{\rule{0.400pt}{1.850pt}}
\multiput(346.60,363.00)(0.468,5.160){5}{\rule{0.113pt}{3.700pt}}
\multiput(345.17,363.00)(4.000,28.320){2}{\rule{0.400pt}{1.850pt}}
\multiput(350.60,399.00)(0.468,5.014){5}{\rule{0.113pt}{3.600pt}}
\multiput(349.17,399.00)(4.000,27.528){2}{\rule{0.400pt}{1.800pt}}
\multiput(354.60,434.00)(0.468,4.868){5}{\rule{0.113pt}{3.500pt}}
\multiput(353.17,434.00)(4.000,26.736){2}{\rule{0.400pt}{1.750pt}}
\multiput(358.60,468.00)(0.468,4.575){5}{\rule{0.113pt}{3.300pt}}
\multiput(357.17,468.00)(4.000,25.151){2}{\rule{0.400pt}{1.650pt}}
\multiput(362.60,500.00)(0.468,4.429){5}{\rule{0.113pt}{3.200pt}}
\multiput(361.17,500.00)(4.000,24.358){2}{\rule{0.400pt}{1.600pt}}
\multiput(366.60,531.00)(0.468,3.990){5}{\rule{0.113pt}{2.900pt}}
\multiput(365.17,531.00)(4.000,21.981){2}{\rule{0.400pt}{1.450pt}}
\multiput(370.59,559.00)(0.477,2.714){7}{\rule{0.115pt}{2.100pt}}
\multiput(369.17,559.00)(5.000,20.641){2}{\rule{0.400pt}{1.050pt}}
\multiput(375.60,584.00)(0.468,3.259){5}{\rule{0.113pt}{2.400pt}}
\multiput(374.17,584.00)(4.000,18.019){2}{\rule{0.400pt}{1.200pt}}
\multiput(379.60,607.00)(0.468,2.821){5}{\rule{0.113pt}{2.100pt}}
\multiput(378.17,607.00)(4.000,15.641){2}{\rule{0.400pt}{1.050pt}}
\multiput(383.60,627.00)(0.468,2.528){5}{\rule{0.113pt}{1.900pt}}
\multiput(382.17,627.00)(4.000,14.056){2}{\rule{0.400pt}{0.950pt}}
\multiput(387.60,645.00)(0.468,2.090){5}{\rule{0.113pt}{1.600pt}}
\multiput(386.17,645.00)(4.000,11.679){2}{\rule{0.400pt}{0.800pt}}
\multiput(391.60,660.00)(0.468,1.651){5}{\rule{0.113pt}{1.300pt}}
\multiput(390.17,660.00)(4.000,9.302){2}{\rule{0.400pt}{0.650pt}}
\multiput(395.60,672.00)(0.468,1.358){5}{\rule{0.113pt}{1.100pt}}
\multiput(394.17,672.00)(4.000,7.717){2}{\rule{0.400pt}{0.550pt}}
\multiput(399.60,682.00)(0.468,1.066){5}{\rule{0.113pt}{0.900pt}}
\multiput(398.17,682.00)(4.000,6.132){2}{\rule{0.400pt}{0.450pt}}
\multiput(403.60,690.00)(0.468,0.774){5}{\rule{0.113pt}{0.700pt}}
\multiput(402.17,690.00)(4.000,4.547){2}{\rule{0.400pt}{0.350pt}}
\multiput(407.00,696.60)(0.481,0.468){5}{\rule{0.500pt}{0.113pt}}
\multiput(407.00,695.17)(2.962,4.000){2}{\rule{0.250pt}{0.400pt}}
\put(411,700.17){\rule{0.900pt}{0.400pt}}
\multiput(411.00,699.17)(2.132,2.000){2}{\rule{0.450pt}{0.400pt}}
\put(415,701.67){\rule{0.964pt}{0.400pt}}
\multiput(415.00,701.17)(2.000,1.000){2}{\rule{0.482pt}{0.400pt}}
\put(419,701.67){\rule{0.964pt}{0.400pt}}
\multiput(419.00,702.17)(2.000,-1.000){2}{\rule{0.482pt}{0.400pt}}
\multiput(423.00,700.95)(0.685,-0.447){3}{\rule{0.633pt}{0.108pt}}
\multiput(423.00,701.17)(2.685,-3.000){2}{\rule{0.317pt}{0.400pt}}
\multiput(427.00,697.95)(0.685,-0.447){3}{\rule{0.633pt}{0.108pt}}
\multiput(427.00,698.17)(2.685,-3.000){2}{\rule{0.317pt}{0.400pt}}
\multiput(431.00,694.93)(0.487,-0.477){7}{\rule{0.500pt}{0.115pt}}
\multiput(431.00,695.17)(3.962,-5.000){2}{\rule{0.250pt}{0.400pt}}
\multiput(436.60,688.51)(0.468,-0.627){5}{\rule{0.113pt}{0.600pt}}
\multiput(435.17,689.75)(4.000,-3.755){2}{\rule{0.400pt}{0.300pt}}
\multiput(440.60,683.09)(0.468,-0.774){5}{\rule{0.113pt}{0.700pt}}
\multiput(439.17,684.55)(4.000,-4.547){2}{\rule{0.400pt}{0.350pt}}
\multiput(444.60,676.68)(0.468,-0.920){5}{\rule{0.113pt}{0.800pt}}
\multiput(443.17,678.34)(4.000,-5.340){2}{\rule{0.400pt}{0.400pt}}
\multiput(448.60,669.26)(0.468,-1.066){5}{\rule{0.113pt}{0.900pt}}
\multiput(447.17,671.13)(4.000,-6.132){2}{\rule{0.400pt}{0.450pt}}
\multiput(452.60,661.68)(0.468,-0.920){5}{\rule{0.113pt}{0.800pt}}
\multiput(451.17,663.34)(4.000,-5.340){2}{\rule{0.400pt}{0.400pt}}
\multiput(456.60,653.85)(0.468,-1.212){5}{\rule{0.113pt}{1.000pt}}
\multiput(455.17,655.92)(4.000,-6.924){2}{\rule{0.400pt}{0.500pt}}
\multiput(460.60,644.85)(0.468,-1.212){5}{\rule{0.113pt}{1.000pt}}
\multiput(459.17,646.92)(4.000,-6.924){2}{\rule{0.400pt}{0.500pt}}
\multiput(464.60,635.85)(0.468,-1.212){5}{\rule{0.113pt}{1.000pt}}
\multiput(463.17,637.92)(4.000,-6.924){2}{\rule{0.400pt}{0.500pt}}
\multiput(468.60,626.85)(0.468,-1.212){5}{\rule{0.113pt}{1.000pt}}
\multiput(467.17,628.92)(4.000,-6.924){2}{\rule{0.400pt}{0.500pt}}
\multiput(472.60,617.85)(0.468,-1.212){5}{\rule{0.113pt}{1.000pt}}
\multiput(471.17,619.92)(4.000,-6.924){2}{\rule{0.400pt}{0.500pt}}
\multiput(476.60,608.43)(0.468,-1.358){5}{\rule{0.113pt}{1.100pt}}
\multiput(475.17,610.72)(4.000,-7.717){2}{\rule{0.400pt}{0.550pt}}
\multiput(480.60,598.43)(0.468,-1.358){5}{\rule{0.113pt}{1.100pt}}
\multiput(479.17,600.72)(4.000,-7.717){2}{\rule{0.400pt}{0.550pt}}
\multiput(484.60,588.85)(0.468,-1.212){5}{\rule{0.113pt}{1.000pt}}
\multiput(483.17,590.92)(4.000,-6.924){2}{\rule{0.400pt}{0.500pt}}
\multiput(488.60,579.43)(0.468,-1.358){5}{\rule{0.113pt}{1.100pt}}
\multiput(487.17,581.72)(4.000,-7.717){2}{\rule{0.400pt}{0.550pt}}
\multiput(492.59,570.26)(0.477,-1.044){7}{\rule{0.115pt}{0.900pt}}
\multiput(491.17,572.13)(5.000,-8.132){2}{\rule{0.400pt}{0.450pt}}
\multiput(497.60,559.85)(0.468,-1.212){5}{\rule{0.113pt}{1.000pt}}
\multiput(496.17,561.92)(4.000,-6.924){2}{\rule{0.400pt}{0.500pt}}
\multiput(501.60,550.43)(0.468,-1.358){5}{\rule{0.113pt}{1.100pt}}
\multiput(500.17,552.72)(4.000,-7.717){2}{\rule{0.400pt}{0.550pt}}
\multiput(505.60,540.85)(0.468,-1.212){5}{\rule{0.113pt}{1.000pt}}
\multiput(504.17,542.92)(4.000,-6.924){2}{\rule{0.400pt}{0.500pt}}
\multiput(509.60,531.43)(0.468,-1.358){5}{\rule{0.113pt}{1.100pt}}
\multiput(508.17,533.72)(4.000,-7.717){2}{\rule{0.400pt}{0.550pt}}
\multiput(513.60,521.85)(0.468,-1.212){5}{\rule{0.113pt}{1.000pt}}
\multiput(512.17,523.92)(4.000,-6.924){2}{\rule{0.400pt}{0.500pt}}
\multiput(517.60,512.85)(0.468,-1.212){5}{\rule{0.113pt}{1.000pt}}
\multiput(516.17,514.92)(4.000,-6.924){2}{\rule{0.400pt}{0.500pt}}
\multiput(521.60,503.85)(0.468,-1.212){5}{\rule{0.113pt}{1.000pt}}
\multiput(520.17,505.92)(4.000,-6.924){2}{\rule{0.400pt}{0.500pt}}
\multiput(525.60,494.85)(0.468,-1.212){5}{\rule{0.113pt}{1.000pt}}
\multiput(524.17,496.92)(4.000,-6.924){2}{\rule{0.400pt}{0.500pt}}
\multiput(529.60,486.26)(0.468,-1.066){5}{\rule{0.113pt}{0.900pt}}
\multiput(528.17,488.13)(4.000,-6.132){2}{\rule{0.400pt}{0.450pt}}
\multiput(533.60,477.85)(0.468,-1.212){5}{\rule{0.113pt}{1.000pt}}
\multiput(532.17,479.92)(4.000,-6.924){2}{\rule{0.400pt}{0.500pt}}
\multiput(537.60,469.26)(0.468,-1.066){5}{\rule{0.113pt}{0.900pt}}
\multiput(536.17,471.13)(4.000,-6.132){2}{\rule{0.400pt}{0.450pt}}
\multiput(541.60,461.26)(0.468,-1.066){5}{\rule{0.113pt}{0.900pt}}
\multiput(540.17,463.13)(4.000,-6.132){2}{\rule{0.400pt}{0.450pt}}
\multiput(545.60,453.26)(0.468,-1.066){5}{\rule{0.113pt}{0.900pt}}
\multiput(544.17,455.13)(4.000,-6.132){2}{\rule{0.400pt}{0.450pt}}
\multiput(549.60,445.26)(0.468,-1.066){5}{\rule{0.113pt}{0.900pt}}
\multiput(548.17,447.13)(4.000,-6.132){2}{\rule{0.400pt}{0.450pt}}
\multiput(553.59,437.93)(0.477,-0.821){7}{\rule{0.115pt}{0.740pt}}
\multiput(552.17,439.46)(5.000,-6.464){2}{\rule{0.400pt}{0.370pt}}
\multiput(558.60,429.68)(0.468,-0.920){5}{\rule{0.113pt}{0.800pt}}
\multiput(557.17,431.34)(4.000,-5.340){2}{\rule{0.400pt}{0.400pt}}
\multiput(562.60,422.26)(0.468,-1.066){5}{\rule{0.113pt}{0.900pt}}
\multiput(561.17,424.13)(4.000,-6.132){2}{\rule{0.400pt}{0.450pt}}
\multiput(566.60,414.68)(0.468,-0.920){5}{\rule{0.113pt}{0.800pt}}
\multiput(565.17,416.34)(4.000,-5.340){2}{\rule{0.400pt}{0.400pt}}
\multiput(570.60,407.68)(0.468,-0.920){5}{\rule{0.113pt}{0.800pt}}
\multiput(569.17,409.34)(4.000,-5.340){2}{\rule{0.400pt}{0.400pt}}
\multiput(574.60,400.68)(0.468,-0.920){5}{\rule{0.113pt}{0.800pt}}
\multiput(573.17,402.34)(4.000,-5.340){2}{\rule{0.400pt}{0.400pt}}
\multiput(578.60,394.09)(0.468,-0.774){5}{\rule{0.113pt}{0.700pt}}
\multiput(577.17,395.55)(4.000,-4.547){2}{\rule{0.400pt}{0.350pt}}
\multiput(582.60,387.68)(0.468,-0.920){5}{\rule{0.113pt}{0.800pt}}
\multiput(581.17,389.34)(4.000,-5.340){2}{\rule{0.400pt}{0.400pt}}
\multiput(586.60,381.09)(0.468,-0.774){5}{\rule{0.113pt}{0.700pt}}
\multiput(585.17,382.55)(4.000,-4.547){2}{\rule{0.400pt}{0.350pt}}
\multiput(590.60,375.09)(0.468,-0.774){5}{\rule{0.113pt}{0.700pt}}
\multiput(589.17,376.55)(4.000,-4.547){2}{\rule{0.400pt}{0.350pt}}
\multiput(594.60,369.09)(0.468,-0.774){5}{\rule{0.113pt}{0.700pt}}
\multiput(593.17,370.55)(4.000,-4.547){2}{\rule{0.400pt}{0.350pt}}
\multiput(598.60,363.09)(0.468,-0.774){5}{\rule{0.113pt}{0.700pt}}
\multiput(597.17,364.55)(4.000,-4.547){2}{\rule{0.400pt}{0.350pt}}
\multiput(602.60,357.09)(0.468,-0.774){5}{\rule{0.113pt}{0.700pt}}
\multiput(601.17,358.55)(4.000,-4.547){2}{\rule{0.400pt}{0.350pt}}
\multiput(606.60,351.51)(0.468,-0.627){5}{\rule{0.113pt}{0.600pt}}
\multiput(605.17,352.75)(4.000,-3.755){2}{\rule{0.400pt}{0.300pt}}
\multiput(610.60,346.09)(0.468,-0.774){5}{\rule{0.113pt}{0.700pt}}
\multiput(609.17,347.55)(4.000,-4.547){2}{\rule{0.400pt}{0.350pt}}
\multiput(614.00,341.93)(0.487,-0.477){7}{\rule{0.500pt}{0.115pt}}
\multiput(614.00,342.17)(3.962,-5.000){2}{\rule{0.250pt}{0.400pt}}
\multiput(619.60,335.51)(0.468,-0.627){5}{\rule{0.113pt}{0.600pt}}
\multiput(618.17,336.75)(4.000,-3.755){2}{\rule{0.400pt}{0.300pt}}
\multiput(623.60,330.51)(0.468,-0.627){5}{\rule{0.113pt}{0.600pt}}
\multiput(622.17,331.75)(4.000,-3.755){2}{\rule{0.400pt}{0.300pt}}
\multiput(627.60,325.51)(0.468,-0.627){5}{\rule{0.113pt}{0.600pt}}
\multiput(626.17,326.75)(4.000,-3.755){2}{\rule{0.400pt}{0.300pt}}
\multiput(631.60,320.51)(0.468,-0.627){5}{\rule{0.113pt}{0.600pt}}
\multiput(630.17,321.75)(4.000,-3.755){2}{\rule{0.400pt}{0.300pt}}
\multiput(635.00,316.94)(0.481,-0.468){5}{\rule{0.500pt}{0.113pt}}
\multiput(635.00,317.17)(2.962,-4.000){2}{\rule{0.250pt}{0.400pt}}
\multiput(639.60,311.51)(0.468,-0.627){5}{\rule{0.113pt}{0.600pt}}
\multiput(638.17,312.75)(4.000,-3.755){2}{\rule{0.400pt}{0.300pt}}
\multiput(643.00,307.94)(0.481,-0.468){5}{\rule{0.500pt}{0.113pt}}
\multiput(643.00,308.17)(2.962,-4.000){2}{\rule{0.250pt}{0.400pt}}
\multiput(647.00,303.94)(0.481,-0.468){5}{\rule{0.500pt}{0.113pt}}
\multiput(647.00,304.17)(2.962,-4.000){2}{\rule{0.250pt}{0.400pt}}
\multiput(651.00,299.94)(0.481,-0.468){5}{\rule{0.500pt}{0.113pt}}
\multiput(651.00,300.17)(2.962,-4.000){2}{\rule{0.250pt}{0.400pt}}
\multiput(655.00,295.94)(0.481,-0.468){5}{\rule{0.500pt}{0.113pt}}
\multiput(655.00,296.17)(2.962,-4.000){2}{\rule{0.250pt}{0.400pt}}
\multiput(659.00,291.94)(0.481,-0.468){5}{\rule{0.500pt}{0.113pt}}
\multiput(659.00,292.17)(2.962,-4.000){2}{\rule{0.250pt}{0.400pt}}
\multiput(663.00,287.94)(0.481,-0.468){5}{\rule{0.500pt}{0.113pt}}
\multiput(663.00,288.17)(2.962,-4.000){2}{\rule{0.250pt}{0.400pt}}
\multiput(667.00,283.94)(0.481,-0.468){5}{\rule{0.500pt}{0.113pt}}
\multiput(667.00,284.17)(2.962,-4.000){2}{\rule{0.250pt}{0.400pt}}
\multiput(671.00,279.95)(0.685,-0.447){3}{\rule{0.633pt}{0.108pt}}
\multiput(671.00,280.17)(2.685,-3.000){2}{\rule{0.317pt}{0.400pt}}
\multiput(675.00,276.94)(0.627,-0.468){5}{\rule{0.600pt}{0.113pt}}
\multiput(675.00,277.17)(3.755,-4.000){2}{\rule{0.300pt}{0.400pt}}
\multiput(680.00,272.95)(0.685,-0.447){3}{\rule{0.633pt}{0.108pt}}
\multiput(680.00,273.17)(2.685,-3.000){2}{\rule{0.317pt}{0.400pt}}
\multiput(684.00,269.94)(0.481,-0.468){5}{\rule{0.500pt}{0.113pt}}
\multiput(684.00,270.17)(2.962,-4.000){2}{\rule{0.250pt}{0.400pt}}
\multiput(688.00,265.95)(0.685,-0.447){3}{\rule{0.633pt}{0.108pt}}
\multiput(688.00,266.17)(2.685,-3.000){2}{\rule{0.317pt}{0.400pt}}
\multiput(692.00,262.95)(0.685,-0.447){3}{\rule{0.633pt}{0.108pt}}
\multiput(692.00,263.17)(2.685,-3.000){2}{\rule{0.317pt}{0.400pt}}
\multiput(696.00,259.95)(0.685,-0.447){3}{\rule{0.633pt}{0.108pt}}
\multiput(696.00,260.17)(2.685,-3.000){2}{\rule{0.317pt}{0.400pt}}
\multiput(700.00,256.95)(0.685,-0.447){3}{\rule{0.633pt}{0.108pt}}
\multiput(700.00,257.17)(2.685,-3.000){2}{\rule{0.317pt}{0.400pt}}
\multiput(704.00,253.95)(0.685,-0.447){3}{\rule{0.633pt}{0.108pt}}
\multiput(704.00,254.17)(2.685,-3.000){2}{\rule{0.317pt}{0.400pt}}
\multiput(708.00,250.95)(0.685,-0.447){3}{\rule{0.633pt}{0.108pt}}
\multiput(708.00,251.17)(2.685,-3.000){2}{\rule{0.317pt}{0.400pt}}
\multiput(712.00,247.95)(0.685,-0.447){3}{\rule{0.633pt}{0.108pt}}
\multiput(712.00,248.17)(2.685,-3.000){2}{\rule{0.317pt}{0.400pt}}
\multiput(716.00,244.95)(0.685,-0.447){3}{\rule{0.633pt}{0.108pt}}
\multiput(716.00,245.17)(2.685,-3.000){2}{\rule{0.317pt}{0.400pt}}
\put(720,241.17){\rule{0.900pt}{0.400pt}}
\multiput(720.00,242.17)(2.132,-2.000){2}{\rule{0.450pt}{0.400pt}}
\multiput(724.00,239.95)(0.685,-0.447){3}{\rule{0.633pt}{0.108pt}}
\multiput(724.00,240.17)(2.685,-3.000){2}{\rule{0.317pt}{0.400pt}}
\put(728,236.17){\rule{0.900pt}{0.400pt}}
\multiput(728.00,237.17)(2.132,-2.000){2}{\rule{0.450pt}{0.400pt}}
\multiput(732.00,234.95)(0.685,-0.447){3}{\rule{0.633pt}{0.108pt}}
\multiput(732.00,235.17)(2.685,-3.000){2}{\rule{0.317pt}{0.400pt}}
\put(736,231.17){\rule{1.100pt}{0.400pt}}
\multiput(736.00,232.17)(2.717,-2.000){2}{\rule{0.550pt}{0.400pt}}
\put(741,229.17){\rule{0.900pt}{0.400pt}}
\multiput(741.00,230.17)(2.132,-2.000){2}{\rule{0.450pt}{0.400pt}}
\multiput(745.00,227.95)(0.685,-0.447){3}{\rule{0.633pt}{0.108pt}}
\multiput(745.00,228.17)(2.685,-3.000){2}{\rule{0.317pt}{0.400pt}}
\put(749,224.17){\rule{0.900pt}{0.400pt}}
\multiput(749.00,225.17)(2.132,-2.000){2}{\rule{0.450pt}{0.400pt}}
\put(753,222.17){\rule{0.900pt}{0.400pt}}
\multiput(753.00,223.17)(2.132,-2.000){2}{\rule{0.450pt}{0.400pt}}
\put(757,220.17){\rule{0.900pt}{0.400pt}}
\multiput(757.00,221.17)(2.132,-2.000){2}{\rule{0.450pt}{0.400pt}}
\put(761,218.17){\rule{0.900pt}{0.400pt}}
\multiput(761.00,219.17)(2.132,-2.000){2}{\rule{0.450pt}{0.400pt}}
\put(765,216.17){\rule{0.900pt}{0.400pt}}
\multiput(765.00,217.17)(2.132,-2.000){2}{\rule{0.450pt}{0.400pt}}
\put(769,214.17){\rule{0.900pt}{0.400pt}}
\multiput(769.00,215.17)(2.132,-2.000){2}{\rule{0.450pt}{0.400pt}}
\put(773,212.17){\rule{0.900pt}{0.400pt}}
\multiput(773.00,213.17)(2.132,-2.000){2}{\rule{0.450pt}{0.400pt}}
\put(777,210.17){\rule{0.900pt}{0.400pt}}
\multiput(777.00,211.17)(2.132,-2.000){2}{\rule{0.450pt}{0.400pt}}
\put(781,208.17){\rule{0.900pt}{0.400pt}}
\multiput(781.00,209.17)(2.132,-2.000){2}{\rule{0.450pt}{0.400pt}}
\put(785,206.17){\rule{0.900pt}{0.400pt}}
\multiput(785.00,207.17)(2.132,-2.000){2}{\rule{0.450pt}{0.400pt}}
\put(789,204.67){\rule{0.964pt}{0.400pt}}
\multiput(789.00,205.17)(2.000,-1.000){2}{\rule{0.482pt}{0.400pt}}
\put(793,203.17){\rule{0.900pt}{0.400pt}}
\multiput(793.00,204.17)(2.132,-2.000){2}{\rule{0.450pt}{0.400pt}}
\put(797,201.17){\rule{1.100pt}{0.400pt}}
\multiput(797.00,202.17)(2.717,-2.000){2}{\rule{0.550pt}{0.400pt}}
\put(802,199.67){\rule{0.964pt}{0.400pt}}
\multiput(802.00,200.17)(2.000,-1.000){2}{\rule{0.482pt}{0.400pt}}
\put(806,198.17){\rule{0.900pt}{0.400pt}}
\multiput(806.00,199.17)(2.132,-2.000){2}{\rule{0.450pt}{0.400pt}}
\put(810,196.67){\rule{0.964pt}{0.400pt}}
\multiput(810.00,197.17)(2.000,-1.000){2}{\rule{0.482pt}{0.400pt}}
\put(814,195.17){\rule{0.900pt}{0.400pt}}
\multiput(814.00,196.17)(2.132,-2.000){2}{\rule{0.450pt}{0.400pt}}
\put(818,193.67){\rule{0.964pt}{0.400pt}}
\multiput(818.00,194.17)(2.000,-1.000){2}{\rule{0.482pt}{0.400pt}}
\put(822,192.17){\rule{0.900pt}{0.400pt}}
\multiput(822.00,193.17)(2.132,-2.000){2}{\rule{0.450pt}{0.400pt}}
\put(826,190.67){\rule{0.964pt}{0.400pt}}
\multiput(826.00,191.17)(2.000,-1.000){2}{\rule{0.482pt}{0.400pt}}
\put(830,189.17){\rule{0.900pt}{0.400pt}}
\multiput(830.00,190.17)(2.132,-2.000){2}{\rule{0.450pt}{0.400pt}}
\put(834,187.67){\rule{0.964pt}{0.400pt}}
\multiput(834.00,188.17)(2.000,-1.000){2}{\rule{0.482pt}{0.400pt}}
\put(838,186.67){\rule{0.964pt}{0.400pt}}
\multiput(838.00,187.17)(2.000,-1.000){2}{\rule{0.482pt}{0.400pt}}
\put(842,185.17){\rule{0.900pt}{0.400pt}}
\multiput(842.00,186.17)(2.132,-2.000){2}{\rule{0.450pt}{0.400pt}}
\put(846,183.67){\rule{0.964pt}{0.400pt}}
\multiput(846.00,184.17)(2.000,-1.000){2}{\rule{0.482pt}{0.400pt}}
\put(850,182.67){\rule{0.964pt}{0.400pt}}
\multiput(850.00,183.17)(2.000,-1.000){2}{\rule{0.482pt}{0.400pt}}
\put(854,181.67){\rule{1.204pt}{0.400pt}}
\multiput(854.00,182.17)(2.500,-1.000){2}{\rule{0.602pt}{0.400pt}}
\put(859,180.67){\rule{0.964pt}{0.400pt}}
\multiput(859.00,181.17)(2.000,-1.000){2}{\rule{0.482pt}{0.400pt}}
\put(863,179.17){\rule{0.900pt}{0.400pt}}
\multiput(863.00,180.17)(2.132,-2.000){2}{\rule{0.450pt}{0.400pt}}
\put(867,177.67){\rule{0.964pt}{0.400pt}}
\multiput(867.00,178.17)(2.000,-1.000){2}{\rule{0.482pt}{0.400pt}}
\put(871,176.67){\rule{0.964pt}{0.400pt}}
\multiput(871.00,177.17)(2.000,-1.000){2}{\rule{0.482pt}{0.400pt}}
\put(875,175.67){\rule{0.964pt}{0.400pt}}
\multiput(875.00,176.17)(2.000,-1.000){2}{\rule{0.482pt}{0.400pt}}
\put(879,174.67){\rule{0.964pt}{0.400pt}}
\multiput(879.00,175.17)(2.000,-1.000){2}{\rule{0.482pt}{0.400pt}}
\put(883,173.67){\rule{0.964pt}{0.400pt}}
\multiput(883.00,174.17)(2.000,-1.000){2}{\rule{0.482pt}{0.400pt}}
\put(887,172.67){\rule{0.964pt}{0.400pt}}
\multiput(887.00,173.17)(2.000,-1.000){2}{\rule{0.482pt}{0.400pt}}
\put(891,171.67){\rule{0.964pt}{0.400pt}}
\multiput(891.00,172.17)(2.000,-1.000){2}{\rule{0.482pt}{0.400pt}}
\put(895,170.67){\rule{0.964pt}{0.400pt}}
\multiput(895.00,171.17)(2.000,-1.000){2}{\rule{0.482pt}{0.400pt}}
\put(899,169.67){\rule{0.964pt}{0.400pt}}
\multiput(899.00,170.17)(2.000,-1.000){2}{\rule{0.482pt}{0.400pt}}
\put(903,168.67){\rule{0.964pt}{0.400pt}}
\multiput(903.00,169.17)(2.000,-1.000){2}{\rule{0.482pt}{0.400pt}}
\put(907,167.67){\rule{0.964pt}{0.400pt}}
\multiput(907.00,168.17)(2.000,-1.000){2}{\rule{0.482pt}{0.400pt}}
\put(911,166.67){\rule{0.964pt}{0.400pt}}
\multiput(911.00,167.17)(2.000,-1.000){2}{\rule{0.482pt}{0.400pt}}
\put(232.0,113.0){\rule[-0.200pt]{13.731pt}{0.400pt}}
\put(920,165.67){\rule{0.964pt}{0.400pt}}
\multiput(920.00,166.17)(2.000,-1.000){2}{\rule{0.482pt}{0.400pt}}
\put(924,164.67){\rule{0.964pt}{0.400pt}}
\multiput(924.00,165.17)(2.000,-1.000){2}{\rule{0.482pt}{0.400pt}}
\put(928,163.67){\rule{0.964pt}{0.400pt}}
\multiput(928.00,164.17)(2.000,-1.000){2}{\rule{0.482pt}{0.400pt}}
\put(932,162.67){\rule{0.964pt}{0.400pt}}
\multiput(932.00,163.17)(2.000,-1.000){2}{\rule{0.482pt}{0.400pt}}
\put(936,161.67){\rule{0.964pt}{0.400pt}}
\multiput(936.00,162.17)(2.000,-1.000){2}{\rule{0.482pt}{0.400pt}}
\put(915.0,167.0){\rule[-0.200pt]{1.204pt}{0.400pt}}
\put(944,160.67){\rule{0.964pt}{0.400pt}}
\multiput(944.00,161.17)(2.000,-1.000){2}{\rule{0.482pt}{0.400pt}}
\put(948,159.67){\rule{0.964pt}{0.400pt}}
\multiput(948.00,160.17)(2.000,-1.000){2}{\rule{0.482pt}{0.400pt}}
\put(952,158.67){\rule{0.964pt}{0.400pt}}
\multiput(952.00,159.17)(2.000,-1.000){2}{\rule{0.482pt}{0.400pt}}
\put(940.0,162.0){\rule[-0.200pt]{0.964pt}{0.400pt}}
\put(960,157.67){\rule{0.964pt}{0.400pt}}
\multiput(960.00,158.17)(2.000,-1.000){2}{\rule{0.482pt}{0.400pt}}
\put(964,156.67){\rule{0.964pt}{0.400pt}}
\multiput(964.00,157.17)(2.000,-1.000){2}{\rule{0.482pt}{0.400pt}}
\put(956.0,159.0){\rule[-0.200pt]{0.964pt}{0.400pt}}
\put(972,155.67){\rule{0.964pt}{0.400pt}}
\multiput(972.00,156.17)(2.000,-1.000){2}{\rule{0.482pt}{0.400pt}}
\put(976,154.67){\rule{1.204pt}{0.400pt}}
\multiput(976.00,155.17)(2.500,-1.000){2}{\rule{0.602pt}{0.400pt}}
\put(968.0,157.0){\rule[-0.200pt]{0.964pt}{0.400pt}}
\put(985,153.67){\rule{0.964pt}{0.400pt}}
\multiput(985.00,154.17)(2.000,-1.000){2}{\rule{0.482pt}{0.400pt}}
\put(981.0,155.0){\rule[-0.200pt]{0.964pt}{0.400pt}}
\put(993,152.67){\rule{0.964pt}{0.400pt}}
\multiput(993.00,153.17)(2.000,-1.000){2}{\rule{0.482pt}{0.400pt}}
\put(997,151.67){\rule{0.964pt}{0.400pt}}
\multiput(997.00,152.17)(2.000,-1.000){2}{\rule{0.482pt}{0.400pt}}
\put(989.0,154.0){\rule[-0.200pt]{0.964pt}{0.400pt}}
\put(1005,150.67){\rule{0.964pt}{0.400pt}}
\multiput(1005.00,151.17)(2.000,-1.000){2}{\rule{0.482pt}{0.400pt}}
\put(1001.0,152.0){\rule[-0.200pt]{0.964pt}{0.400pt}}
\put(1013,149.67){\rule{0.964pt}{0.400pt}}
\multiput(1013.00,150.17)(2.000,-1.000){2}{\rule{0.482pt}{0.400pt}}
\put(1009.0,151.0){\rule[-0.200pt]{0.964pt}{0.400pt}}
\put(1021,148.67){\rule{0.964pt}{0.400pt}}
\multiput(1021.00,149.17)(2.000,-1.000){2}{\rule{0.482pt}{0.400pt}}
\put(1017.0,150.0){\rule[-0.200pt]{0.964pt}{0.400pt}}
\put(1029,147.67){\rule{0.964pt}{0.400pt}}
\multiput(1029.00,148.17)(2.000,-1.000){2}{\rule{0.482pt}{0.400pt}}
\put(1025.0,149.0){\rule[-0.200pt]{0.964pt}{0.400pt}}
\put(1037,146.67){\rule{1.204pt}{0.400pt}}
\multiput(1037.00,147.17)(2.500,-1.000){2}{\rule{0.602pt}{0.400pt}}
\put(1033.0,148.0){\rule[-0.200pt]{0.964pt}{0.400pt}}
\put(1046,145.67){\rule{0.964pt}{0.400pt}}
\multiput(1046.00,146.17)(2.000,-1.000){2}{\rule{0.482pt}{0.400pt}}
\put(1042.0,147.0){\rule[-0.200pt]{0.964pt}{0.400pt}}
\put(1054,144.67){\rule{0.964pt}{0.400pt}}
\multiput(1054.00,145.17)(2.000,-1.000){2}{\rule{0.482pt}{0.400pt}}
\put(1050.0,146.0){\rule[-0.200pt]{0.964pt}{0.400pt}}
\put(1062,143.67){\rule{0.964pt}{0.400pt}}
\multiput(1062.00,144.17)(2.000,-1.000){2}{\rule{0.482pt}{0.400pt}}
\put(1058.0,145.0){\rule[-0.200pt]{0.964pt}{0.400pt}}
\put(1074,142.67){\rule{0.964pt}{0.400pt}}
\multiput(1074.00,143.17)(2.000,-1.000){2}{\rule{0.482pt}{0.400pt}}
\put(1066.0,144.0){\rule[-0.200pt]{1.927pt}{0.400pt}}
\put(1082,141.67){\rule{0.964pt}{0.400pt}}
\multiput(1082.00,142.17)(2.000,-1.000){2}{\rule{0.482pt}{0.400pt}}
\put(1078.0,143.0){\rule[-0.200pt]{0.964pt}{0.400pt}}
\put(1094,140.67){\rule{0.964pt}{0.400pt}}
\multiput(1094.00,141.17)(2.000,-1.000){2}{\rule{0.482pt}{0.400pt}}
\put(1086.0,142.0){\rule[-0.200pt]{1.927pt}{0.400pt}}
\put(1107,139.67){\rule{0.964pt}{0.400pt}}
\multiput(1107.00,140.17)(2.000,-1.000){2}{\rule{0.482pt}{0.400pt}}
\put(1098.0,141.0){\rule[-0.200pt]{2.168pt}{0.400pt}}
\put(1115,138.67){\rule{0.964pt}{0.400pt}}
\multiput(1115.00,139.17)(2.000,-1.000){2}{\rule{0.482pt}{0.400pt}}
\put(1111.0,140.0){\rule[-0.200pt]{0.964pt}{0.400pt}}
\put(1127,137.67){\rule{0.964pt}{0.400pt}}
\multiput(1127.00,138.17)(2.000,-1.000){2}{\rule{0.482pt}{0.400pt}}
\put(1119.0,139.0){\rule[-0.200pt]{1.927pt}{0.400pt}}
\put(1143,136.67){\rule{0.964pt}{0.400pt}}
\multiput(1143.00,137.17)(2.000,-1.000){2}{\rule{0.482pt}{0.400pt}}
\put(1131.0,138.0){\rule[-0.200pt]{2.891pt}{0.400pt}}
\put(1155,135.67){\rule{0.964pt}{0.400pt}}
\multiput(1155.00,136.17)(2.000,-1.000){2}{\rule{0.482pt}{0.400pt}}
\put(1147.0,137.0){\rule[-0.200pt]{1.927pt}{0.400pt}}
\put(1168,134.67){\rule{0.964pt}{0.400pt}}
\multiput(1168.00,135.17)(2.000,-1.000){2}{\rule{0.482pt}{0.400pt}}
\put(1159.0,136.0){\rule[-0.200pt]{2.168pt}{0.400pt}}
\put(1184,133.67){\rule{0.964pt}{0.400pt}}
\multiput(1184.00,134.17)(2.000,-1.000){2}{\rule{0.482pt}{0.400pt}}
\put(1172.0,135.0){\rule[-0.200pt]{2.891pt}{0.400pt}}
\put(1200,132.67){\rule{0.964pt}{0.400pt}}
\multiput(1200.00,133.17)(2.000,-1.000){2}{\rule{0.482pt}{0.400pt}}
\put(1188.0,134.0){\rule[-0.200pt]{2.891pt}{0.400pt}}
\put(1216,131.67){\rule{0.964pt}{0.400pt}}
\multiput(1216.00,132.17)(2.000,-1.000){2}{\rule{0.482pt}{0.400pt}}
\put(1204.0,133.0){\rule[-0.200pt]{2.891pt}{0.400pt}}
\put(1237,130.67){\rule{0.964pt}{0.400pt}}
\multiput(1237.00,131.17)(2.000,-1.000){2}{\rule{0.482pt}{0.400pt}}
\put(1220.0,132.0){\rule[-0.200pt]{4.095pt}{0.400pt}}
\put(1257,129.67){\rule{0.964pt}{0.400pt}}
\multiput(1257.00,130.17)(2.000,-1.000){2}{\rule{0.482pt}{0.400pt}}
\put(1241.0,131.0){\rule[-0.200pt]{3.854pt}{0.400pt}}
\put(1277,128.67){\rule{0.964pt}{0.400pt}}
\multiput(1277.00,129.17)(2.000,-1.000){2}{\rule{0.482pt}{0.400pt}}
\put(1261.0,130.0){\rule[-0.200pt]{3.854pt}{0.400pt}}
\put(1302,127.67){\rule{0.964pt}{0.400pt}}
\multiput(1302.00,128.17)(2.000,-1.000){2}{\rule{0.482pt}{0.400pt}}
\put(1281.0,129.0){\rule[-0.200pt]{5.059pt}{0.400pt}}
\put(1326,126.67){\rule{0.964pt}{0.400pt}}
\multiput(1326.00,127.17)(2.000,-1.000){2}{\rule{0.482pt}{0.400pt}}
\put(1306.0,128.0){\rule[-0.200pt]{4.818pt}{0.400pt}}
\put(1355,125.67){\rule{0.964pt}{0.400pt}}
\multiput(1355.00,126.17)(2.000,-1.000){2}{\rule{0.482pt}{0.400pt}}
\put(1330.0,127.0){\rule[-0.200pt]{6.022pt}{0.400pt}}
\put(1383,124.67){\rule{0.964pt}{0.400pt}}
\multiput(1383.00,125.17)(2.000,-1.000){2}{\rule{0.482pt}{0.400pt}}
\put(1359.0,126.0){\rule[-0.200pt]{5.782pt}{0.400pt}}
\put(1416,123.67){\rule{0.964pt}{0.400pt}}
\multiput(1416.00,124.17)(2.000,-1.000){2}{\rule{0.482pt}{0.400pt}}
\put(1387.0,125.0){\rule[-0.200pt]{6.986pt}{0.400pt}}
\put(1420.0,124.0){\rule[-0.200pt]{3.854pt}{0.400pt}}
\end{picture}

\vskip 1.0 truein
\caption{ Plot of $P(\Omega, z)$ from (26) at redshift = $z_R$ as
a function of $\delta = 1-\Omega$.
 This probability distribution comes from (24) assuming $L = C_R$
and is intended for illustrative purposes only. The effects of
additional redshifting are indicated by the factors $\times (1+z)^{\pm 2}$.
}
\end{center}
\end{figure}

\end{document}